# New Parametrizations of Non-Gaussian Line-of-sight Velocity Distribution

HongSheng Zhao

Max-Planck-Institute für Astrophysik, 85740 Garching, Germany

Email: hsz@MPA-Garching.MPG.DE

Francisco Prada

Max-Planck-Institute für Extraterrestrische Physik, 85740 Garching, Germany

Instituto de Astrofisica de Canarias, 38200, La Laguna, Tenerife, Spain[1]

Email: fprada@ll.iac.es

## ABSTRACT

A five-parameter fitting formula for the line-of-sight stellar velocity distributions of steady state systems is proposed. It can faithfully reproduce velocity distributions of theoretical models ranging from nearly Gaussian profiles to strongly skewed or mildly double-peaked profiles. In contrast to van der Marel and Franx (1993) and Kuijken and Merrifield (1993), the line profiles are required to have neither multi-peaks nor negative velocity wings and most informations can be recovered from no more than five physically meaningful fitting parameters.

*Subject headings*: galaxies: kinematics and dynamics - line: profiles

---

[1] Present address

## 1. Introduction

Exactly Gaussian stellar velocity distributions exist only in simplified theoretical models. Line-of-sight velocity distributions (LOSVDs) of realistic stellar systems, which are implied by broadened spectral features in a galaxy spectrum by the Doppler effect, commonly show deviations from Gaussian in terms of a non-zero skewness and kurtosis, and a fair fraction of them can be slightly double-peaked. When fitting broadened spectral lines from observation, it is important to parametrize the LOSVD more generally than a three-parameter Gaussian because the deviations from Gaussian carry as much critical information of the stellar dynamics of the underlying system as the line-strength, the mean velocity and the dispersion. For example, LOSVDs have been used to identify kinematically distinct cores in some elliptical galaxies (Franx and Illingworth 1988, Rix and White 1992) and set non-trivial limit on the dark halo in an E0 galaxy (Carollo et al. 1995). Similarly in two spiral galaxies analysis of LOSVD have revealed that a fraction of stars in the disc are counter-rotating (Rix et al. 1992, Merrifield and Kuijken 1994). Another recent application is the detection of a bar in two edge-on spiral galaxies with boxy bulges based on the presence of double-peaked stellar LOSVD (Kuijken and Merrifield 1995).

When fitting a typical galaxy spectrum in the presence of noise and systematic fitting errors, it is likely that the well-determined parameters is only a handful. It appears to be more sensible to fit the observation with a simply parametrized velocity profile than a profile with many parameters or a fully non-parametric profile. If all realistic profiles were exactly Gaussian, then the number of meaningful parameters would be only three, namely, the amplitude, the mean velocity and the dispersion. If realistic profiles were genericly very close to Gaussian, then the number of meaningful parameters would be no more than five; the two additional parameters would be the skewness and the kurtosis. In this paper we show that even for profiles with large deviation from Gaussian can still be specified by no more than five parameters.

Realistic profiles are likely much more restricted than an arbitrary positive function of velocity, because the velocity and density distributions of a steady state system satisfy several equations including the Vlasov equation and the Poisson equation. Previous methods are often designed to cover all smooth positive and negative mathematical functions in a complete basis set as the Gauss-Hermite expansion method by Gerhard (1993) and van der Marel and Franx (1993, hereafter vdMF), or smooth positive functions as the unresolved Gaussian decomposition method by Kuijken and Merrifield (1993, hereafter KM). These parametrizations also tolerate multiple peaks, peaks at the wings and even negative wings in the profile, which are likely spurious due to noise and artifacts of the parametrizations. Many free parameters are often invoked to fit theoretical profiles or observed spectrum, typically about 5 to 9 parameters for vdMF, and on the order of 20 parameters for KM.

Different from these methods, our philosophy is to maximally exploit the *a priori* information of realistic profiles so as to represent a general but realistic profile with only a few parameters. The parametrization is tailored to model the majority of realistic line profiles which have the following properties by assumption.

(I) They are positive and smooth everywhere in velocity.

(II) They have no more than three points with vanishing first derivative excluding at infinity, namely, they are either single-peaked profiles with up to two flat shoulders or double-peaked profiles.

(III) The deviation from a Gaussian at the wings is much smaller than the peak amplitude and/or is not significantly above the noise level.

These conditions are fairly general, and are often met by line profiles of theoretical models and of observed systems (see references in the beginning of the Introduction) where any strong deviation from Gaussian is near the systematic velocity. They also include most of the interesting classes of line profiles from slightly skewed profiles to mildly double-peaked profiles.

In this paper we give parametrizations with the above properties, and compare with previous methods. Both Gerhard (1993) and vdMF use the Gaussian-Hermite expansion parametrization; the former uses a fixed Gaussian as the zero order term while the latter uses the best fitting Gaussian. Our comparison will be mostly with the more related vdMF method.

The paper is organized as following. §2 gives the parametrization and its properties. It is then tested with synthesized profiles from theoretical models in §3. The conclusion and a brief discussion of possible generalizations are given in §4.

## 2. The Parametrization

### 2.1. A modified Gaussian-Hermite parametrization

Motivated by the fact that Gaussian is still a fair approximation to realistic profiles, we write a general profile $L(v)$ as the following,

$$L(v) = \frac{\gamma}{\sqrt{2\pi}\sigma} e^{-\frac{w^2}{2}} [1 + Q(w)] \tag{1}$$

where
$$w = \frac{v - V}{\sigma}, \quad (2)$$
is the rescaled velocity, $\gamma$, $V$, $\sigma$ are the line strength, mean velocity and dispersion of the best fitting Gaussian. $Q(w)$ specifies the amount of deviation from the Gaussian.

In vdMF $Q(w)$ is given by an expansion on the complete Gaussian-Hermite function set $H_n(w)$ with $n \geq 3$ (see their equation 9). Namely,
$$Q_{vdMF}(w) = \sum_{n=3}^{N} c_n H_n(w), \quad (3)$$
where $c_n$ is a set of fitting coefficients for the amplitude of deviations from Gaussian, and $H_n(w)$ is the n-th Gaussian-Hermite function defined in vdMF, which satisfies
$$\int_{-\infty}^{+\infty} H_m(y) H_n(y) e^{-y^2} dy = \sqrt{\pi} \delta_{mn}, \quad (4)$$
and are more specificly given by
$$H_3(y) = \frac{2}{\sqrt{3}}(y^3 - \frac{3}{2}y), \quad (5)$$
and
$$H_4(y) = \frac{2}{\sqrt{6}}(y^4 - 2y^2 + \frac{3}{4}). \quad (6)$$

$Q_{vdMF}(w)$ is essentially a polynomial function of $w$ of order three or higher, which like virtually all polynomial functions is wildly oscillating and asymptoticly reaches positive or negative infinity. An example is shown by the dashed lines (labeled $b = 0$) in Figure 1 for (up to a constant factor) $H_4(w)$ (right panel) and $H_3(w)$ (left panel). Due to these behaviours of $Q(w)$, the line profile $L(v)$ often has multiple peaks or negative wings or artificial peaks at the wings.

For realistic profiles with the three properties given in the Introduction, we propose to modify the vdMF parametrization and to regularize $Q(w)$ and $L(v)$ at the wings by introducing Gaussian damping terms $\exp(-b_n w^2/2)$ with the constants $b_n > 0$. Generally we write
$$Q(w) = \sum_{n=3}^{N} c_n a_n e^{-\frac{b_n w^2}{2}} H_n(a_n w), \quad a_n = \sqrt{1 + \frac{b_n}{2}}. \quad (7)$$
This parametrization is designed so that it reduces to $Q_{vdMF}(w)$ if $b_n = 0$ for $n = 1, N$. In practice we find it is sufficient to truncate the expansion at $N = 4$. As a result the total number of fitting parameters are only five, namely, $(\gamma, V, \sigma, c_3, c_4)$. We further set the constants $b_3 = b_4 = b$ and impose one of the three cases of damping at the wings: zero damping ($b = 0$), medium damping ($b = 1$) and strong damping ($b = 2$); the zero damping corresponds to vdMF and merely serves as a reference case.

Specificly the proposed parametrization of line profiles is as follows,

$$L(v) = \frac{\gamma}{\sqrt{2\pi}\sigma} e^{-\frac{w^2}{2}}[1 + Q(w)] \tag{8}$$

$$= \frac{1}{\sqrt{2\pi}\sigma} e^{-\frac{w^2}{2}} \{\gamma + a e^{-\frac{bw^2}{2}} [\gamma_3 H_3(aw) + \gamma_4 H_4(aw)]\} \tag{9}$$

where

$$b \subset \{0, 1, 2\}, \quad a \equiv \sqrt{1 + \frac{b}{2}}, \quad \gamma_3 \equiv c_3 \gamma, \quad \gamma_4 \equiv c_4 \gamma, \quad w = \frac{v - V}{\sigma}. \tag{10}$$

### 2.2. Properties similar to Gaussian-Hermite expansions

Similar to vdMF, the parameters $(\gamma, V, \sigma, c_3, c_4)$ used here are distinct moments of the line profile $L(v)$. They are related to $L(v)$ by the following equations.

$$\gamma = \sqrt{2} \int_{-\infty}^{+\infty} L(v) e^{-\frac{w^2}{2}} dv, \tag{11}$$

$$V = \frac{1}{\gamma} \int_{-\infty}^{+\infty} v L(v) e^{-\frac{w^2}{2}} dv, \tag{12}$$

$$\sigma = [\frac{1}{\gamma} \int_{-\infty}^{+\infty} (v - V)^2 L(v) e^{-\frac{w^2}{2}} dv]^{\frac{1}{2}}, \tag{13}$$

$$c_3 = \frac{\sqrt{2}}{\gamma k_3} \int_{-\infty}^{+\infty} L(v) e^{-\frac{(1+b)w^2}{2}} H_3(aw) dv, \tag{14}$$

$$c_4 = \frac{\sqrt{2}}{\gamma k_4} \int_{-\infty}^{+\infty} L(v) e^{-\frac{(1+b)w^2}{2}} H_4(aw) dv. \tag{15}$$

The constants $k_3$ and $k_4$ are given in the Appendix. These relations can be easily shown if one applies equation (4) and also notes that cross terms between an even and an odd Gaussian-Hermite function are always zero and that by design

$$\int_{-\infty}^{+\infty} H_m(w) Q(w) e^{-w^2} dw = 0 \text{ for } m = 0, 1, 2. \tag{16}$$

The above equations imply that our parametrization as in equations (9)-(10) preserves the nice properties of vdMF. First $(\gamma, V, \sigma)$ are the parameters for the best fitting Gaussian. The parameters $c_3$ and $c_4$ describe the anti-symmetric and the symmetric deviations from Gaussian; they reduce to $h_3$ and $h_4$ if $b = 0$ or if the deviation from Gaussian is small.

Second the basis set in our parametrization is an orthogonal basis set and the parameters correlate only weakly (due to non-linearity). Fixing $V$ and $\sigma$ so that the parametrization is a linear function of the fitting parameters $(\gamma, \gamma_3, \gamma_4)$, the correlation matrix of $(\gamma, \gamma_3, \gamma_4)$ is diagonal and equals to the identity matrix. The nearly diagonal covariance matrix simplifies the error analysis.

Third it is also easy to implement the convolution with template spectrum with our parametrization. The Fourier transformation of $L(v)$ is simplely

$$\int_{-\infty}^{+\infty} L(v) e^{ifv} dv = \gamma e^{-\frac{\sigma^2}{2}f^2} + e^{-\frac{\sigma^2}{2(1+b)}f^2} \{-i\gamma_3 q^{-\frac{7}{2}} [H_3(\frac{\sigma}{a}f) + \sqrt{3} r^2] \quad (17)$$

$$+ \gamma_4 q^{-\frac{9}{2}} [H_4(\frac{\sigma}{a}f) + \sqrt{12} r^2 H_2(\frac{\sigma}{a}f) + \sqrt{\frac{3}{2}} r^4]\}$$

where

$$q = 1 + r = 1 + \frac{b}{2+b}, \quad a = \sqrt{1 + \frac{b}{2}}. \quad (18)$$

### 2.3. New properties

The main benefit of the modification is that the spurious positive or negative peaks at the wings in vdMF can be effectively suppressed in this parametrization. With damping $Q(w)$ is finite everywhere and asymptoticly approaches zero because mathematically the Gaussian damping term dominate the polynomials at large velocities. Hence for sufficiently small $c_n$, $1 + Q(w)$ is positive everywhere and unity at infinity. The resulting profile $L(v)$ can deviate from Gaussian only near the systematic velocity, not at the wings. The behaviours of $Q(w)$ in two cases $b = 1$ and $b = 2$ as well as the $b = 0$ reference case are illustrated by the solid lines in Figure 1, which show (up to a constant factor) $H_n(w\sqrt{1+b/2})\exp(-bw^2/2)$ versus $w$ for $n = 3$ (right panel) and $n = 4$ (left panel).

Fig. 2 shows some typical profiles in each parametrization. While vdMF (as shown by the $b = 0$ undamped case) genericly contains negative wings which is significant even for slightly flat-topped profiles or skewed profiles, the damped cases ($b = 1$ or $b = 2$) show no negative wings. The $b = 1$ profiles range from nearly Gaussian to mildly double-peaked profiles, including flat-topped and strongly skewed profiles. The $b = 2$ profiles can also be strongly double-peaked. Still we prefer the parametrization for the $b = 1$ case with equation (9)-(10) because strongly two-peaked profiles are relatively rare compared to skewed and mildly double-peaked profiles. In general the profiles with $b = 1$ also approaches to Gaussian at the wings slower than the $b = 2$ profiles, hence can incorporate non-Gaussian wings, e.g., a power-law fall-off, near $2.5\sigma$.

The shape of our profiles is completely specified by $c_3$ and $c_4$. Fig. 3 and Fig. 4 show the regions of $c_3 - c_4$ that prescribe smooth positive profiles with one, or two, or three peaks for the $b = 1$ case and for the $b = 2$ case. When $c_3$ and $c_4$ is small (approximately $0 < c_4 < 0.3$ and $|c_3 \pm c_4| < 0.3$), the profile is single peaked and $(\gamma, V, \sigma, c_3, c_4)$ are approximately the amplitude, the mean velocity, the dispersion, and up to a constant (see Appendix) the skewness and the kurtosis of the profile. For sufficiently big $c_3$ or $c_4$, the profile will contain negative wings or multiple ($\geq 3$) peaks. For intermediate $c_3$ and $c_4$, the profile becomes double-peaked. The meaning of the parameters are less direct than, e.g., the ratio and the offset between the two peaks. In these cases one can use Table I (if $b = 1$) and Table II (if $b = 2$) to convert $c_3$ and $c_4$ to more direct characteristic shape parameters of the profile and scale the physical dimensions with $(\gamma, V, \sigma)$.

Comparing with previous parametrizations, our parametrization as in equations (9)-(10) not only can model a wide range of realistic profiles, but also has many nice properties, including using the least number of free parameters. All information of the line profile is contained in five numbers. Because damping makes it possible to fit a range of realistic profiles without introducing spurious wings, there is little need to include higher order Gaussian-Hermite terms such as $H_5(w)$ or $H_6(w)$ in our parametrization. By contrast, one often needs to keep higher order expansion terms $H_N(w)$ in vdMF method with $N + 1 = 5$ to 9 parameters to hide the negative wings effectively and to fit double-peaked profiles. The multi-Gaussian decompositions (KM) and non-parametric fitting effectively invoke even more parameters, on the order of the number of velocity bins. The double-Gaussian decomposition covers similar range of profiles as our parametrization, but the former invokes six parameters, which often do not have clear physical meanings and are often strongly correlated.

## 3. Test with synthesized profiles

In this section we apply the parametrization to some synthesized profiles from simple theoretical models of spheroid and disc systems.

Figure 5 and Figure 6 show a few line profiles of some analytical spheroidal models inside a flattened logarithmic dark halo potential by Evans (1994). These models are flattened and have power-law density profiles and anisotropic velocity ellipsoids. When viewed edge-on on the major axis, the projected line-of-sight velocity distribution of a non-rotating anisotropic model can be written analytically as following.

$$L(v) = e^{-\frac{pv^2}{2}} + \delta e^{-\frac{p+2}{2}v^2}(1 + kv^2), \quad \delta = \frac{s^{-2} - 1}{p}(\frac{p+2}{p})^{\frac{1}{2}}, \quad k = (p-1)(p+2)\frac{R^2}{1+R^2}, \quad (19)$$

where the dark halo core radius and terminal velocity have been set to unity, $p$ is the power-law slope of the stellar spheroid, $s$ in Evans's notation or $\delta$ are the anisotropy constant, $R$ is the major axis distance. Since a rigorous expression for rotating systems is complicated, we simulate effects of rotation qualitatively by replacing the first term $e^{-\frac{pv^2}{2}}$ with $e^{-\frac{p(v-t)^2}{2}}$, where $t$ is a rotation measure.

We have computed a few profiles at different locations ($R = 1, 2, \infty$) in the analytical models with different amount of anisotropy ($\delta = 0.2, 1, 2$), rotation ($t = 0, 0.2, 0.4$) and power-law slope ($p = 3$ for Figure 5 and $p = 4$ for Figure 6) based on equation (19). These profiles are then fitted using both the vdMF method (the heavy dashed line) and our method with $b = 1$ (the solid line). We find that our parametrization fits the analytical profiles better than vdMF in all above cases although larger residuals of our fit start to show up outside the core of the steeper power-law case (Figure 6).

The analytical profiles cover a wide range in shape. First the deviation from Gaussian becomes smaller closer to the core. Second at fixed position $R$ the profiles change continuously from a double-peak structure to a nearly Gaussian shape for decreasing values of anisotropy $\delta$ and/or power-law slope $p$. These profiles are similar in variety to those of other analytical spheroidal models, e.g., those shown in Gerhard (1993). It is also likely that these profiles span the range of profiles of real bulges or ellipticals, which are hot and mildly flattened and anisotropic systems with typical power-law slope less than or equal to 4. While some line profiles of systems with very big anisotropy, flattening and power-law slope are not well-fitted by our parametrization as shown in Figure 6, these systems may be intrinsicly rare in nature due to formation processes and constraints from equilibrium and stability.

It is necessary to caution that the two-integral models used as test cases here are likely too simple for real elliptical galaxies, where the third integral always plays a role. Based on the observed line profiles of a sample of 50 or so ellipticals Bender, Saglia and Gerhard (1994) found that they deviate from a Gaussian by no more than 10%, which would argue that real profiles are typically single-peaked. Deviations are more prominent in the asymmetric part than the symmetric part with $|c_3| \approx |h_3| \leq 0.15$ and $|c_4| \approx |h_4| \leq 0.05$. Profiles with a strong flat-top ($c_4 \approx h_4 < -0.05$) are absent in their sample, which they take as evidences against two-integral flattened non-rotating models being the right dynamical model.

We have also tested the parametrization with simple models of disc systems. We synthesize line profiles of a disc plus bulge systems by superimposing a wider Gaussian with a shifted narrow Gaussian. Similarly we synthesize line profiles of counter-rotating disc systems by superimposing two well-separated Gaussian profiles with equal dispersion.

In both cases we vary the ratio of the two Gaussians to simulate profiles at different major axis distance. Figure 7 shows the fit to these double-Gaussian line profiles. The results are consistent with the finding with analytical spheroid models. Except for some extreme cases as shown in panel 4 where the amplitude ratio is very small, most line profiles are well fit with our parametrization.

## 4. Conclusion and Possible Generalizations

In summary, we describe a method to derive line profiles without spurious wings. The method is based on modifying the Gaussian-Hermite expansion method by Gerhard (1993) and vdMF. The main application of our method is to recover the commonly seen strongly skewed, flat-topped or weakly double-peaked profiles. The power of the proposed parametrization remains to be tested in fitting observed spectrum.

It is possible to generalize the proposed parametrization. In particular the lowest order term needs not be a Gaussian. Many symmetric functions which fall off steeply at large velocities would also serve. The following parametrization, for example, is found to work very well.

$$L(v) = sech\frac{w^2}{2}[1 + (s_1 w' + s_2 w'^2)] \tag{20}$$

where $s_1$ and $s_2$ are fitting coefficients, and

$$w = \frac{v - V}{\sigma}, \ w' = \frac{w}{1 + \frac{w^2}{2}}. \tag{21}$$

Fig. 8 shows fits to a theoretical line profile by five different parametrizations, and their residuals. The theoretical profile is the symmetric double peaked profile shown in Figure 3d of Evans (1994), which is the profile at one core radius for a non-rotating Evans model with the power-law slope $p = s^{-2} = 3.5$ and $\delta = 0.9$. Clearly the vdMF method including $H_4(y)$ (the dashed line) or both $H_4(y)$ and $H_6(y)$ (the heavy dashed line) cannot fit the double peaks and have significant negative wings. The rest three parametrizations fit roughly equally well with the double-Gaussian (the dotted line) and the Sech-parametrization (the faint solid line) being even better than our preferred parametrization (the heavy solid line) in this case. Also curiously our parametrization recovers the true dispersion and kurtosis of the symmetric line profile better than both the double-Gaussian parametrization and the Sech-parametrization as shown in Table 3; the truncated vdMF expansions are also disfavored in this criteria. In general we perfer the parametrization in equation (9)-(10) because it is easy to implement and because the involved parameters are few and uncorrelated.

HSZ acknowledges helpful discussions with Hans-Walter Rix. FP would like to thank the financial support by the E.C. ANTARES network (ERB 4050/PL930536) under contract CHRX-CT 930359.

---

## 5. Appendix

In this appendix we give detailed expressions for a few quantities used in the main text.

(1) The constants $k_3$ and $k_4$ in equations (14) to (15) are given by the following,

$$k_3 = \frac{1}{\sqrt{\pi}} \int_{-\infty}^{+\infty} e^{-(2a^2-1)w^2} [H_3(aw)]^2 a\, dw \tag{1}$$

$$= (1+b)^{-7/2}(1+b/2)^{3/2}(1+b+5b^2/8) \tag{2}$$

$$= 1 \text{ for } b = 0, \tag{3}$$

$$= \frac{63}{256}\sqrt{3} \text{ for } b = 1, \tag{4}$$

$$= \frac{11}{81}\sqrt{6} \text{ for } b = 2, \tag{5}$$

$$k_4 = \frac{1}{\sqrt{\pi}} \int_{-\infty}^{+\infty} e^{-(2a^2-1)w^2} [H_4(aw)]^2 a\, dw \tag{6}$$

$$= (1+b)^{-9/2}(1+b/2)^{1/2}(1+2b+9b^2/4+5b^3/4+35b^4/128) \tag{7}$$

$$= 1 \text{ for } b = 0, \tag{8}$$

$$= \frac{867}{4096}\sqrt{3} \text{ for } b = 1, \tag{9}$$

$$= \frac{227}{1944}\sqrt{6} \text{ for } b = 2, \tag{10}$$

where $a = \sqrt{1+b/2}$.

(2) The following are the asymptotic expressions for the line strength, the mean velocity, the dispersion, the skewness and the kurtosis of our parametrized profile in the limit that the deviation from Gaussian is small.

$$\text{Line strength } \tilde{\gamma} \equiv \int_{-\infty}^{+\infty} L(v) dv \tag{11}$$

$$= \gamma[1 + \frac{(1+b/2)^{1/2}}{4(1+b)^{5/2}}\sqrt{6}c_4] \tag{12}$$

$$= \gamma(1 + \frac{1}{4}\sqrt{6}c_4) \text{ for } b = 0 \tag{13}$$

$$= \gamma(1 + \frac{3}{32}\sqrt{2}c_4) \text{ for } b = 1 \tag{14}$$

$$= \gamma(1 + \frac{1}{18}c_4) \text{ for } b = 2 \tag{15}$$

$$\text{Mean velocity } \tilde{V} \equiv \int_{-\infty}^{+\infty} vL(v)/\tilde{\gamma}\, dv \tag{16}$$

$$\quad = \quad \sigma \frac{(1+b/2)}{(1+b)^{5/2}} \sqrt{3} c_3 \qquad (17)$$

$$= \quad \sigma \sqrt{3} c_3 \text{ for } b = 0 \qquad (18)$$

$$= \quad \sigma \frac{3}{16} \sqrt{6} c_3 \text{ for } b = 1 \qquad (19)$$

$$= \quad \sigma \frac{2}{9} c_3 \text{ for } b = 2 \qquad (20)$$

$$\text{Dispersion } \tilde{\sigma} \equiv [\int_{-\infty}^{+\infty} (v - \tilde{V})^2 L(v)/\tilde{\gamma} dv]^{1/2} \qquad (21)$$

$$= \quad \sigma[1 + \frac{(1+3b/8)(1+b/2)^{1/2}}{(1+b)^{7/2}} \sqrt{6} c_4] \qquad (22)$$

$$= \quad \sigma(1 + \sqrt{6} c_4) \text{ for } b = 0 \qquad (23)$$

$$= \quad \sigma(1 + \frac{33}{128}\sqrt{2} c_4) \text{ for } b = 1 \qquad (24)$$

$$= \quad \sigma(1 + \frac{7}{54} c_4) \text{ for } b = 2 \qquad (25)$$

$$\text{Skewness} \equiv \int_{-\infty}^{+\infty} (v - \tilde{V})^3 L(v)/(\tilde{\gamma}\tilde{\sigma}^3) dv \qquad (26)$$

$$= \quad \frac{(1-b/4)(1+b/2)}{(1+b)^{7/2}} 4\sqrt{3} c_3 \qquad (27)$$

$$= \quad 4\sqrt{3} c_3 \text{ for } b = 0 \qquad (28)$$

$$= \quad \frac{9}{32}\sqrt{6} c_3 \text{ for } b = 1 \qquad (29)$$

$$= \quad \frac{4}{27} c_3 \text{ for } b = 2 \qquad (30)$$

$$\text{Kurtosis} \equiv \int_{-\infty}^{+\infty} (v - \tilde{V})^4 L(v)/(\tilde{\gamma}\tilde{\sigma}^4) dv - 3 \qquad (31)$$

$$= \quad \frac{(1 - b/2 - 13b^2/32)(1+b/2)^{1/2}}{(1+b)^{9/2}} 8\sqrt{6} c_4 \qquad (32)$$

$$= \quad 8\sqrt{6} c_4 \text{ for } b = 0 \qquad (33)$$

$$= \quad \frac{9}{128}\sqrt{2} c_4 \text{ for } b = 1 \qquad (34)$$

$$= \quad -\frac{26}{81} c_4 \text{ for } b = 2 \qquad (35)$$

Table 1. Characteristic parameters of a single or double peaked profile on the grid of $c_3 - c_4$ in the $b = 1$ case

| $c_3, c_4$ | $\frac{P_2}{P_1}, \frac{V_1-V_2}{\sigma}$ | $\frac{D}{P_1}, \frac{V_1-V_D}{\sigma}$ | $P_1, V_1$ | Mean, Disp | Skew, Kurt |
|---|---|---|---|---|---|
| 0.00 , 0.00 | 0.00 , 0.00 | 0.00 , 0.00 | 1.00 , 0.00 | 0.00 , 1.00 | 0.00 , -0.01 |
| 0.00 , 0.10 | 0.00 , 0.00 | 0.00 , 0.00 | 1.08 , 0.00 | 0.00 , 1.03 | 0.00 , -0.02 |
| 0.00 , 0.20 | 0.00 , 0.00 | 0.00 , 0.00 | 1.15 , 0.00 | 0.00 , 1.07 | 0.00 , -0.04 |
| 0.00 , 0.30 | 0.00 , 0.00 | 0.00 , 0.00 | 1.23 , 0.00 | 0.00 , 1.10 | -0.00 , -0.08 |
| 0.10 , 0.00 | 0.00 , 0.20 | 0.00 , 0.00 | 1.03 , 0.00 | 0.05 , 1.00 | 0.07 , -0.03 |
| 0.10 , 0.10 | 0.00 , 0.12 | 0.00 , 0.00 | 1.09 , 0.00 | 0.05 , 1.03 | 0.05 , -0.03 |
| 0.10 , 0.20 | 0.00 , 0.08 | 0.00 , 0.00 | 1.16 , 0.00 | 0.04 , 1.07 | 0.04 , -0.05 |
| 0.10 , 0.30 | 0.36 , 1.44 | 0.36 , 0.28 | 1.23 , 1.36 | 0.04 , 1.10 | 0.03 , -0.08 |
| 0.20 , 0.00 | 0.00 , 0.28 | 0.00 , 0.00 | 1.09 , 0.00 | 0.09 , 1.00 | 0.14 , -0.07 |
| 0.20 , 0.10 | 0.00 , 0.20 | 0.00 , 0.00 | 1.13 , 0.00 | 0.09 , 1.03 | 0.11 , -0.05 |
| 0.20 , 0.20 | 0.42 , 1.40 | 0.42 , 0.24 | 1.19 , 1.24 | 0.09 , 1.06 | 0.08 , -0.07 |
| 0.20 , 0.30 | 0.40 , 1.52 | 0.35 , 0.48 | 1.26 , 1.40 | 0.09 , 1.10 | 0.06 , -0.10 |
| 0.20 , 0.40 | 0.39 , 1.56 | 0.28 , 0.56 | 1.32 , 1.48 | 0.09 , 1.13 | 0.04 , -0.14 |
| 0.30 , 0.00 | 0.53 , 1.24 | 0.53 , 0.08 | 1.15 , 0.92 | 0.14 , 0.99 | 0.22 , -0.13 |
| 0.30 , 0.10 | 0.47 , 1.44 | 0.46 , 0.32 | 1.19 , 1.16 | 0.14 , 1.03 | 0.17 , -0.10 |
| 0.30 , 0.20 | 0.44 , 1.52 | 0.39 , 0.48 | 1.23 , 1.32 | 0.13 , 1.06 | 0.13 , -0.11 |
| 0.30 , 0.30 | 0.42 , 1.60 | 0.32 , 0.60 | 1.29 , 1.44 | 0.13 , 1.09 | 0.09 , -0.13 |
| 0.30 , 0.40 | 0.42 , 1.60 | 0.26 , 0.64 | 1.35 , 1.48 | 0.13 , 1.12 | 0.07 , -0.16 |
| 0.40 , 0.00 | 0.51 , 1.52 | 0.45 , 0.52 | 1.22 , 1.16 | 0.18 , 0.98 | 0.30 , -0.24 |
| 0.40 , 0.10 | 0.48 , 1.60 | 0.40 , 0.56 | 1.25 , 1.28 | 0.18 , 1.02 | 0.23 , -0.18 |
| 0.40 , 0.20 | 0.46 , 1.60 | 0.34 , 0.60 | 1.29 , 1.36 | 0.18 , 1.05 | 0.17 , -0.16 |
| 0.40 , 0.30 | 0.45 , 1.64 | 0.28 , 0.68 | 1.34 , 1.44 | 0.18 , 1.09 | 0.13 , -0.17 |
| 0.40 , 0.40 | 0.44 , 1.64 | 0.22 , 0.68 | 1.39 , 1.48 | 0.17 , 1.12 | 0.09 , -0.19 |
| 0.50 , 0.20 | 0.47 , 1.68 | 0.29 , 0.68 | 1.35 , 1.40 | 0.22 , 1.04 | 0.23 , -0.24 |
| 0.50 , 0.30 | 0.47 , 1.68 | 0.24 , 0.72 | 1.39 , 1.44 | 0.22 , 1.08 | 0.17 , -0.22 |
| 0.50 , 0.40 | 0.46 , 1.68 | 0.19 , 0.72 | 1.44 , 1.48 | 0.22 , 1.11 | 0.12 , -0.23 |

Note. — For each pair of $c_3, c_4$, the table gives the peak ratios and offset between the second peak with the main peak ($\frac{P_2}{P_1}$ and $\frac{V_1-V_2}{\sigma}$), the dip with the main peak ($\frac{D}{P_1}$ and $\frac{V_1-V_D}{\sigma}$), the main peak with the best fitting Gaussian ($P_1$, $V_1$), the mean velocity, the dispersion, the skewness and the kurtosis of the profile.

Table 2.  Same as Table I, but for $b = 2$

| $c_3, c_4$ | $\frac{P_2}{P_1}, \frac{V_1-V_2}{\sigma}$ | $\frac{D}{P_1}, \frac{V_1-V_D}{\sigma}$ | $P_1, V_1$ | Mean, Disp | Skew, Kurt |
|---|---|---|---|---|---|
| 0.00 , -0.50 | 1.00 , 1.28 | 0.50 , 0.64 | 1.14 , 0.64 | 0.00 , 0.93 | 0.00 , 0.13 |
| 0.00 , -0.30 | 1.00 , 1.20 | 0.73 , 0.60 | 1.02 , 0.60 | 0.00 , 0.96 | 0.00 , 0.08 |
| 0.00 , -0.10 | 1.00 , 0.72 | 0.98 , 0.36 | 0.93 , 0.36 | 0.00 , 0.99 | 0.00 , 0.02 |
| 0.00 , 0.10 | 0.00 , 0.00 | 0.00 , 0.00 | 1.09 , 0.00 | 0.00 , 1.01 | 0.00 , -0.04 |
| 0.00 , 0.30 | 0.00 , 0.00 | 0.00 , 0.00 | 1.26 , 0.00 | 0.00 , 1.04 | -0.00 , -0.11 |
| 0.20 , -0.30 | 0.78 , 1.20 | 0.59 , 0.52 | 1.16 , 0.68 | 0.05 , 0.96 | 0.05 , 0.07 |
| 0.20 , -0.10 | 0.70 , 1.08 | 0.69 , 0.24 | 1.10 , 0.68 | 0.04 , 0.99 | 0.04 , 0.01 |
| 0.20 , 0.10 | 0.54 , 1.08 | 0.54 , 0.08 | 1.16 , 0.88 | 0.04 , 1.01 | 0.03 , -0.05 |
| 0.20 , 0.30 | 0.46 , 1.28 | 0.38 , 0.44 | 1.30 , 1.16 | 0.04 , 1.04 | 0.02 , -0.11 |
| 0.40 , -0.10 | 0.59 , 1.32 | 0.46 , 0.52 | 1.28 , 0.92 | 0.09 , 0.98 | 0.07 , -0.01 |
| 0.40 , 0.10 | 0.53 , 1.36 | 0.40 , 0.52 | 1.30 , 1.08 | 0.09 , 1.01 | 0.05 , -0.06 |
| 0.40 , 0.30 | 0.49 , 1.40 | 0.29 , 0.56 | 1.40 , 1.20 | 0.09 , 1.03 | 0.04 , -0.12 |
| 0.60 , 0.10 | 0.52 , 1.44 | 0.25 , 0.64 | 1.46 , 1.12 | 0.13 , 1.00 | 0.08 , -0.09 |
| 0.60 , 0.30 | 0.50 , 1.44 | 0.18 , 0.64 | 1.53 , 1.20 | 0.13 , 1.03 | 0.06 , -0.14 |

Table 3.  Fitted parameters for a symmetric line profile of Evans model by different methods.

|  | Dispersion | Kurtosis |
|---|---|---|
| Evans model | 0.60 | -0.14 |
| $b = 1$ | 0.62 | -0.10 |
| Sech | 0.59 | -0.50 |
| 2-Gaussian | 0.58 | -0.59 |
| vdMF, $N = 4$ | 0.45 | -11.78 |
| vdMF, $N = 6$ | 0.65 | 1.94 |

Note. — compares the dispersion and kurtosis of a line profile from the Evans model with those derived from different methods.

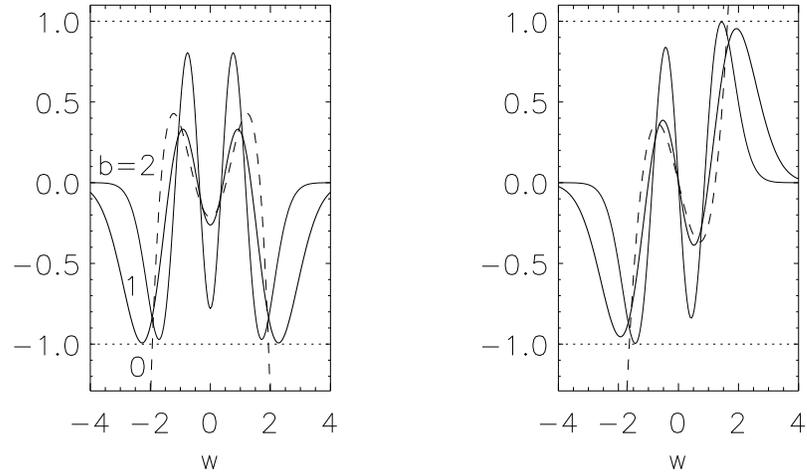

Fig. 1.— The two panels show the fourth order symmetric and the third order anti-symmetric Gauss-Hermite polynomials with different amount of damping $b = 0, 1, 2$. The dashed line (labeled $b = 0$) corresponds to the no-damping case as in vdMF. More specificly the curves show (up to a constant factor) $H_n(w\sqrt{1 + b/2}) \exp(-bw^2/2)$ versus $w$ for $n = 4$ (left panel) and $n = 3$ (right panel).

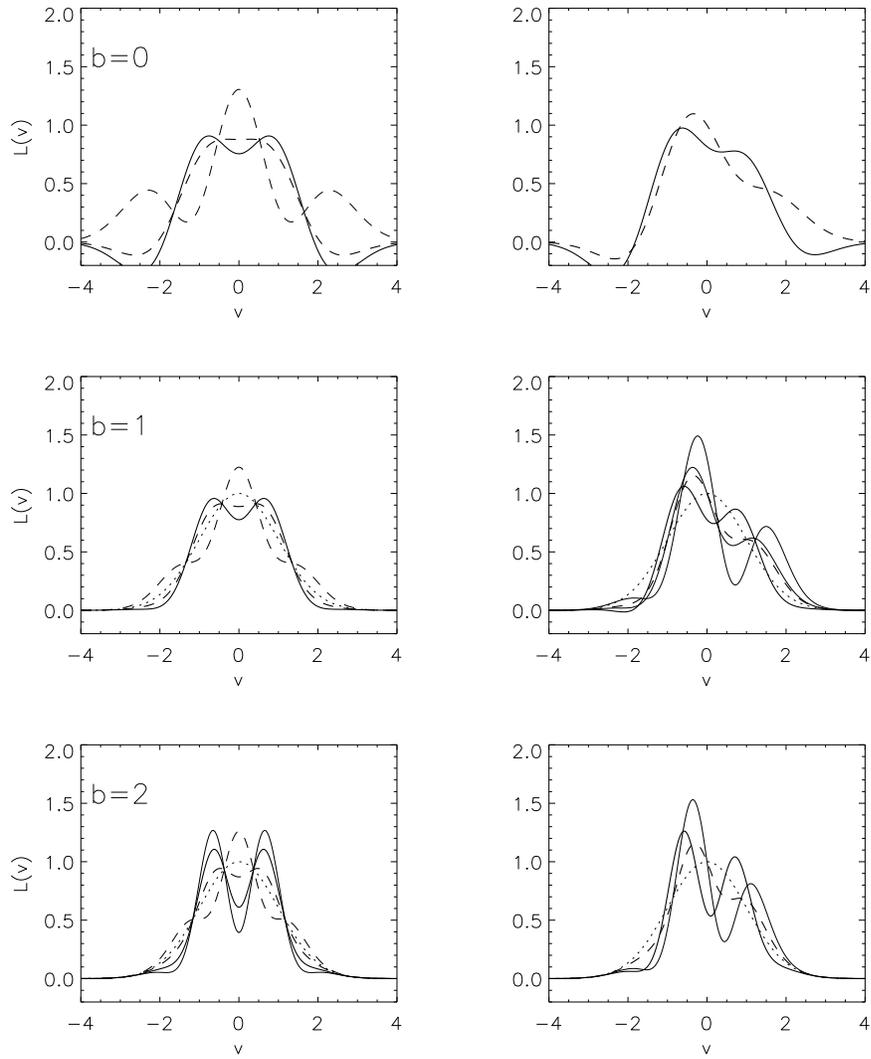

Fig. 2.— The upper, the middle and the lower panels show some typical symmetric and asymmetric profiles for $b = 0, 1, 2$. The solid lines are double-peaked profiles, the dashed lines are strongly squared or sharply peaked or skewed profiles.

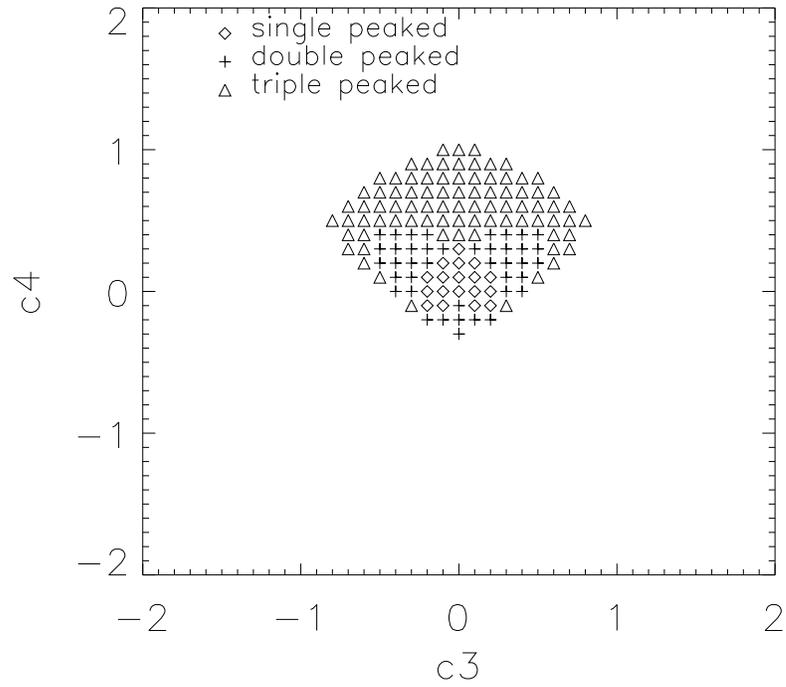

Fig. 3.— shows the regions of $c_3 - c_4$ plane which would yield positive profiles of one, two and three peaks. The damping parameter $b = 1$.

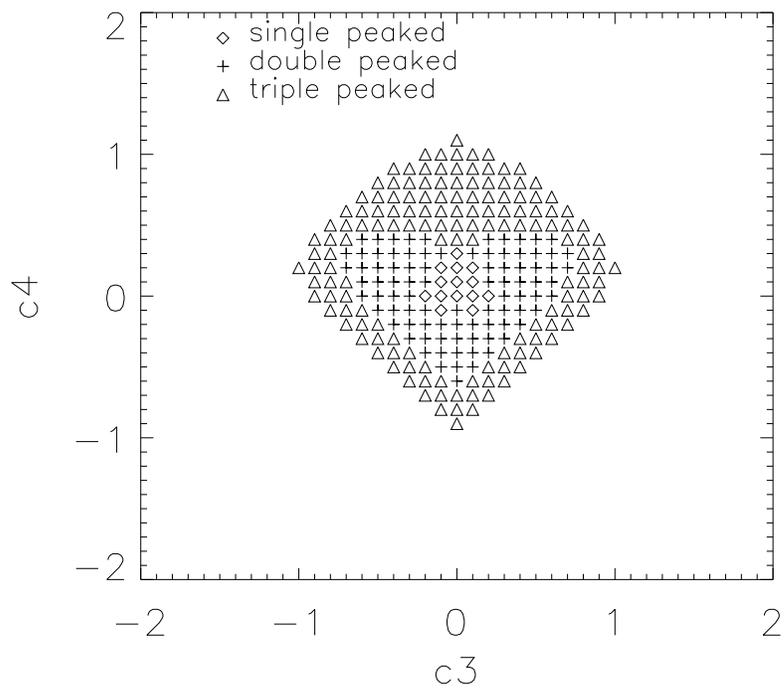

Fig. 4.— same as the previous figure, but for the damping parameter $b = 2$.

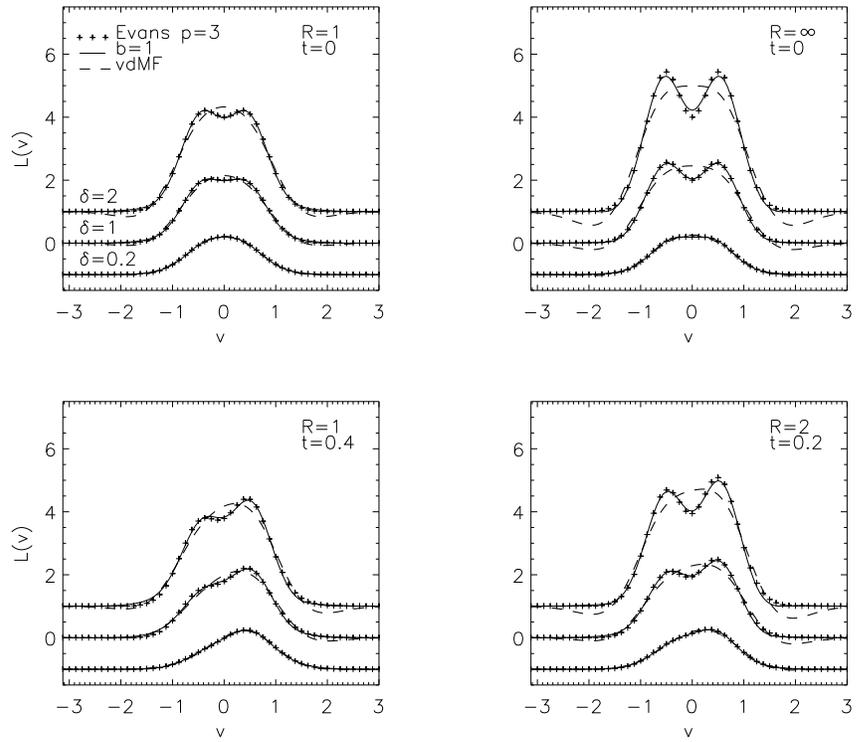

Fig. 5.— shows the fits to line profiles of a simple power-law galaxy model by Evans (1994) with the power-law slope $p = 3$ (the plus symbols) using our method with $b = 1$ (solid lines) and vdMF method with $b = 0$ (dashed lines). The profiles are at different radius $R$ and with different amount of rotation $t$ and anisotropy $\delta$.

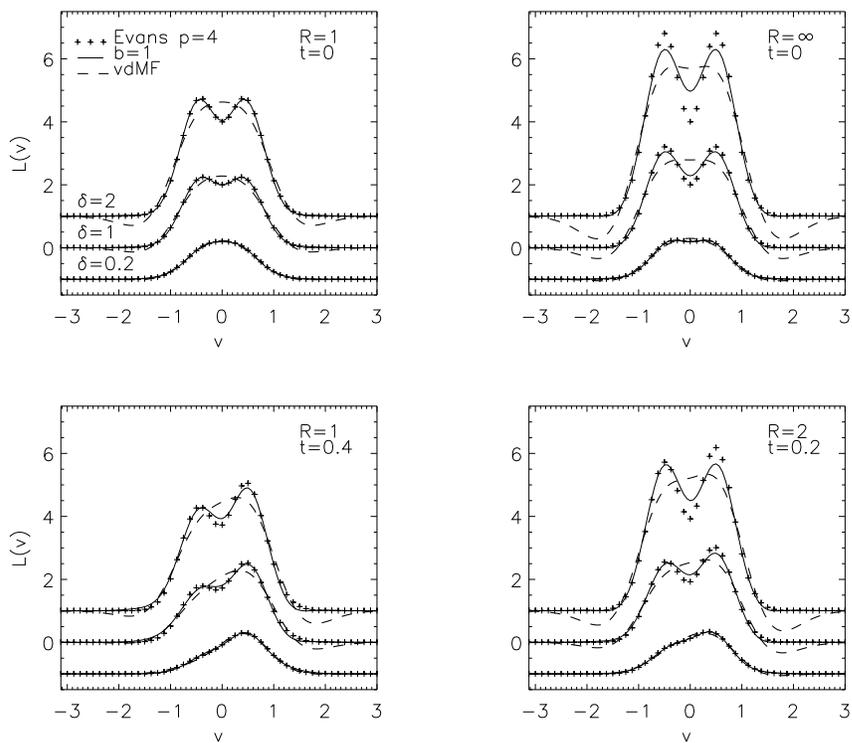

Fig. 6.— same as the previous figure, but for the power-law slope $p = 4$.

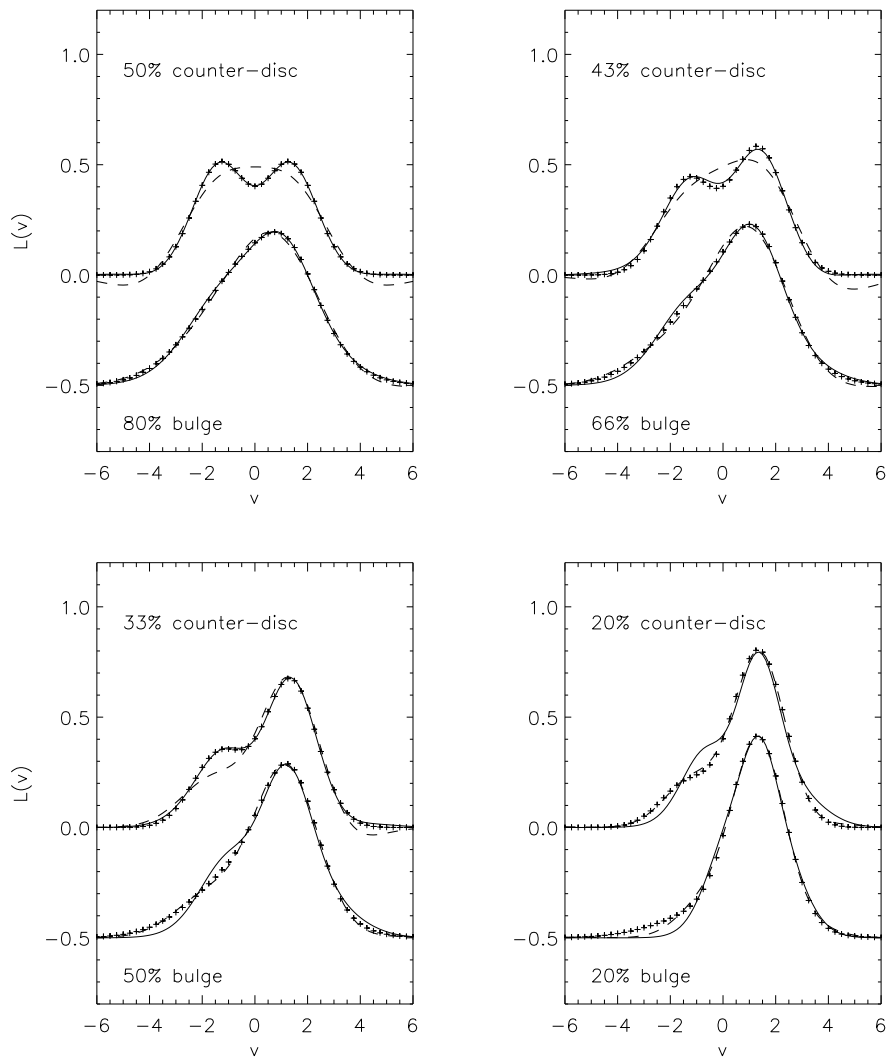

Fig. 7.— shows the fits to double-Gaussian line profiles (the plus symbols) using our method with $b = 1$ (solid lines) and vdMF method with $b = 0$ (dashed lines). Each panel shows one case with two counter-rotating Gaussian discs ($\sigma_{disc} = 1$ and $V_{rot} = \pm 1.35$) and one case with the direct disc plus a non-rotating Gaussian bulge with $\sigma_{bulge} = 2$. The percentage of either the counter-rotating disc or the non-rotating bulge is as labeled.

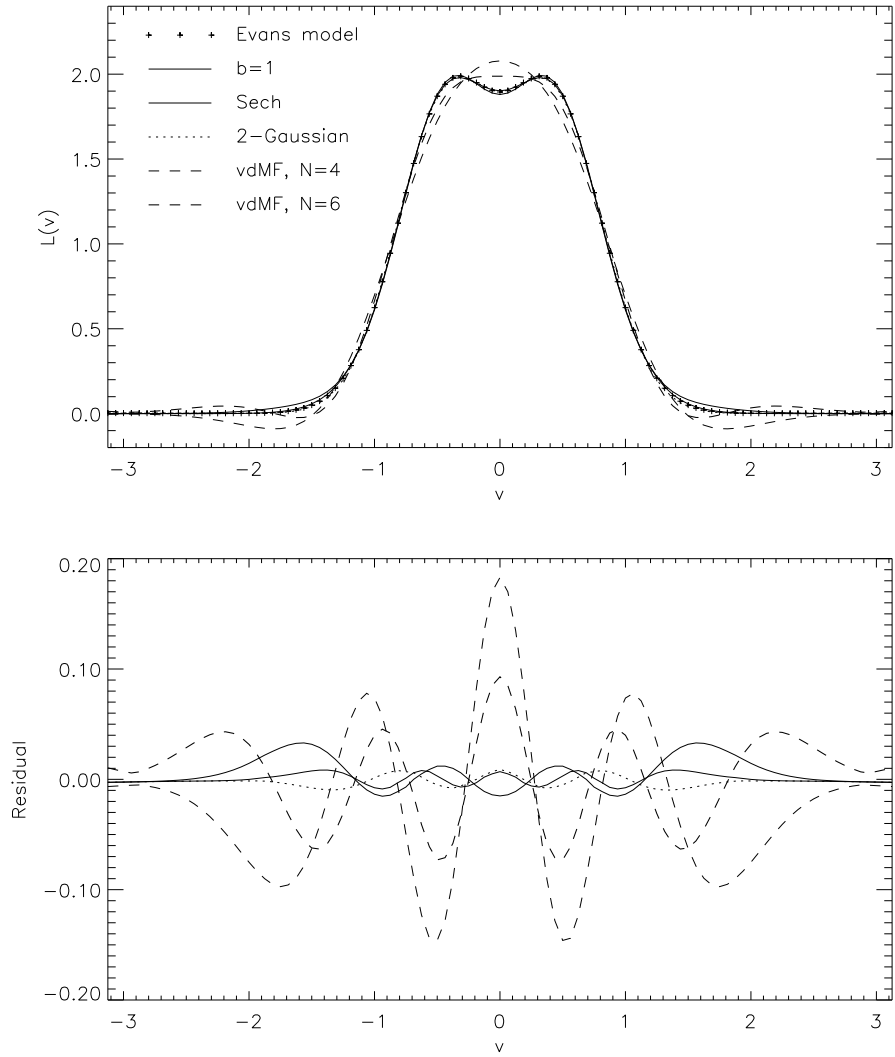

Fig. 8.— shows the fits to a symmetric double-peaked profile from Evans' model (the plus symbols) by five different parametrizations. The heavy solid line and the solid line show our preferred parametrization as in equations (9)-(10) with $b = 1$ and an alternative parametrization as in equation (20). The dashed line and the heavy dashed line are fits with vdMF method as in equations (1), (2) and (3) including up to $H_4(y)$ and $H_6(y)$ respectively. The dotted line shows a fit by double-Gaussian parametrization.